\documentclass[twocolumn,preprintnumbers,amsmath,amssymb,superscriptaddress,prb]{revtex4}

\usepackage{graphicx}
\usepackage{dcolumn}
\usepackage{bm}

\begin{document}


\title{The making of ferromagnetic Fe doped ZnO nano-clusters}

\author{Nirmal Ganguli}
\affiliation{Department of Physics, Indian Institute of Technology Bombay, Mumbai 400076, India}
\affiliation{Department of Solid State Physics and Center for Advanced Materials, Indian Association for the Cultivation of Science, Jadavpur, Kolkata 700032, India}
\author{Indra Dasgupta}
\email[Author to whom any correspondence should be addressed. Email: ]{sspid@iacs.res.in}
\affiliation{Department of Physics, Indian Institute of Technology Bombay, Mumbai 400076, India}
\affiliation{Department of Solid State Physics and Center for Advanced Materials, Indian Association for the Cultivation of Science, Jadavpur, Kolkata 700032, India}
\author{Biplab Sanyal}
\affiliation{Department of Physics and Materials Science, Uppsala University, SE-75121 Uppsala, Sweden}

\date{\today}

\begin{abstract}
In this letter, the authors present a study of the energetics and magnetic interactions in Fe doped ZnO clusters by {\em ab-initio} density functional calculations. The results indicate that defects under suitable conditions can induce ferromagnetic interactions between the dopant Fe atoms whereas antiferromagnetic coupling dominates in a neutral defect-free cluster. The calculations also reveal an unusual ionic state of the dopant Fe atom residing at the surface of the cluster, a feature that is important to render the cluster ferromagnetic.
\end{abstract}

\maketitle

Mn doped GaAs is a prototype diluted magnetic semiconductor (DMS). However its Curie temperature could be raised only to a maximum of $\sim$170 K \cite{jungwirth} making it hardly useful for room temperature spintronic applications. For room temperature ferromagnetism (FM) ZnO and GaN were predicted to be promising as host materials.\cite{dietl} Following this theoretical prediction there has been an avalanche of experimental work to realize room temperature FM in these systems. Among these materials transition metal (TM) doped ZnO received special attention as it was possible to dope TM's not only in bulk ZnO but also in thin films as well as nano-crystalline form of ZnO. However the experimental results on the realization of room temperature FM are highly controversial but there is general consensus that FM strongly depends on methods and conditions employed in the preparation of the samples. 
\begin{figure}
\includegraphics[scale=0.35]{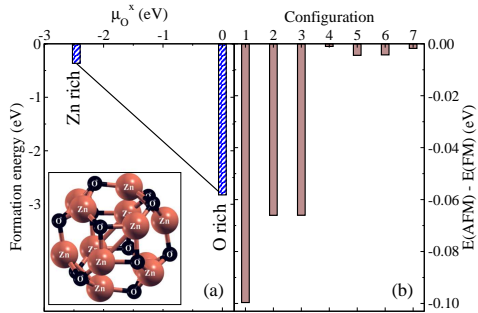}
\caption{\label{fig1}(Color online) (a): Formation energy of {\em one} Fe atom doped at a Zn
site of a ZnO cluster as a function of excess chemical potential of O
($\mu_{o}^{x}$), (b): Energy difference ($\Delta E$) between
antiparallel (AFM) and parallel (FM) magnetic orientations when {\em
two} Fe atoms are doped at the Zn sites as a function of various
configurations (with increasing distance between two Fe
ions in the cluster). Inset: Relaxed structure of the pure ${\rm Zn_{12}O_{12}}$
cluster.}
\end{figure}

Nanoclusters of ZnO capable of sustaining ferromagnetic order are of particular interest since a clear understanding 
of the finite size effect on the magnetic properties of such systems is essential for the development of high-density storage media with nano sized constituent particles or crystallites. In contrast to bulk the nano particles are dominated by surfaces and are therefore naturally susceptible to defects that may have crucial impact on their magnetic properties and future application in a spintronic device. Kittilstved {\em et al.} \cite{kittilstved} have shown the importance of defects for ferromagnetism in Mn and Co-doped ZnO nano-crystals.  Very recently, Karmakar {\em et al.} \cite{karmakar} have indicated that the coexistence of Fe$^{2+}$ and Fe$^{3+}$ states promotes ferromagnetism in Fe doped ZnO nano-cluster. 
They argued that the surface electronic structure of these nanoparticles is quite different from the core presumably due to defects and is dominated by Fe$^{3+}$ ions, which is crucial for the magnetic properties. These observations are further supported by recent x-ray magnetic circular dichroism (XMCD) studies on the same samples.\cite{kataoka}
In the present letter, using {\em ab-initio} electronic structure calculations we have investigated the conditions that will render Fe doped ZnO clusters to be ferromagnetic.

For the electronic structure calculations we have used the projector augmented wave (PAW) method \cite{blochl} as implemented in the VASP package.\cite{vasp,vasp2} PAW potentials with 12 valence electrons for Zn ($3d^{10}4s^2$), 6 for O ($2p^42s^2$) and 8 for Fe ($3d^64s^2$) with an energy cutoff of 500 eV for the plane wave expansion of the PAW's was employed in our calculations. The exchange-correlation part was approximated in the generalized gradient approximation\cite{pbe} including a Hubbard U for the dopant Fe. For Fe, U (on-site $d-d$ Coulomb interaction) = 4.0 eV and J (on-site exchange interaction) = 1.0 eV was chosen.
The formation energy (FE) of the defects was calculated using the following expression:\cite{ossicini,mahadevan,chris}
\begin{eqnarray}
{\rm FE}&=&E{\rm (Zn_mO_n}(\alpha, q)) - E{\rm (Zn_pO_p(pure))} \nonumber \\
&+& \sum_{\alpha} n_{\alpha}\mu_{\alpha} + q(E_v + \epsilon_F)
\label{eq:fe}
\end{eqnarray}
where $\alpha$ is the defect atom added or removed from the pristine ZnO, $n_{\alpha}$ is the number of each 
defect atoms: $n_{\alpha} = -1~(+1)$ for adding (removing) {\em one} atom, 
$E{\rm (Zn_mO_n}(\alpha, q))$ is the total energy with defect 
$\alpha$ and charge $q$, while $E{\rm (Zn_pO_p(pure))}$ is the total 
energy of the pure ZnO cluster. $\epsilon_F$ is the 
Fermi level measured with respect 
to the energy ($E_v$)
of the highest occupied molecular orbital (HOMO) in the pristine 
ZnO cluster.  $\mu_{\alpha}$ is the chemical potential of atom $\alpha$ in some suitable reservoir.  Chemical potentials of Zn and O are obtained 
from the condition: $p(\mu^s_{Zn} + \mu^x_{Zn}+\mu^s_O+\mu^x_O) = E({\rm Zn_pO_p(pure)})$,
where $\mu^s_{Zn}$ and $\mu^s_O$ are the energies per atom of Zn and O respectively in the standard elemental structure and $\mu^x_{Zn}$ and $\mu^x_O$ are excess chemical potentials of Zn and O which obey the above constraint at equilibrium. In addition, the excess chemical potentials should satisfy the relation $\mu^x_{Zn}$ $\le$ 0 and $\mu^x_O$ $\le$ 0 to assure that Zn and O in the elemental structure are not formed. Depending on growth condition, the excess chemical potentials 
can have two extreme limits either $\mu^x_{Zn} = 0$ or $\mu^x_O = 0$ referred to as the Zn rich and O rich limits respectively. 
For Fe we have taken the chemical potential to be the energy per atom in the most commonly occurring bcc structure and assumed $\mu^x_{Fe} = 0$. The total energies of the charged simulation cell were computed by compensating any additional charge by a uniform jellium background.

In view of the recent reports \cite{liu, reber, yadav} 
that ${\rm Zn_{12}O_{12}}$ is the magic cluster with the
 most stable configuration 
we confined our study to the 24-atom ${\rm Zn_{12}O_{12}}$ cluster. The cluster was simulated in a large cell (a cube of 20~\AA) surrounded by vacuum and periodic boundary conditions.  In order to obtain the optimized structure we 
considered different initial structures and performed structural relaxations
which led to the optimized structure displayed in Fig.~1(see Inset).
The structure is spherical with all the atoms on the surface of the sphere providing an unique opportunity to study the crucial surface effects.  All the atoms are very symmetrically arranged with six  4 atom and eight 6 atom rings. The diameter of the cluster is $\sim$6.35~\AA.
The HOMO-LUMO gap for this cluster was found to be 2.34~eV.
\begin{figure}
\includegraphics[scale=0.35]{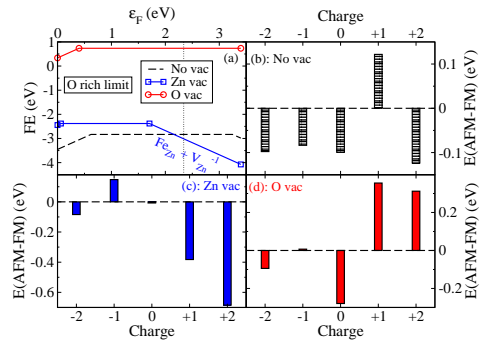}
\caption{\label{fig2}(Color online) (a): Formation energies for {\em one} Fe atom doped at a Zn site in the cluster in the oxygen rich limit with and without vacancies for different charged states. Only the lowest formation energy of a particular charged state is shown. When a vacancy is present, the Fe ion and the vacancy are at the closest possible positions. 
The dotted vertical line indicates the calculated band gap for the cluster; (b), (c), (d): corresponds to $\Delta E$ between AFM and FM magnetic orientation
when a pair of Fe atoms at the closest possible position are doped at the Zn sites
in presence of no vacancy, Zn vacancy, and O vacancy respectively.}
\end{figure}

One atom of the cluster was substituted with an Fe atom and this cluster was fully relaxed. We find that substitution of Zn atom by Fe is energetically more favorable than substitution of an O atom in the ZnO cluster. The formation energy of doping one Fe atom at the Zn site in the cluster has been calculated and is plotted in Fig.~1(a) as a function of excess chemical potential of Oxygen ($\mu_O^x$). As expected, the formation energy of doping Fe at the Zn site is higher in the Zn-rich limit. The singly doped cluster is magnetic with a  magnetic moment of 4~$\mu_B$, as expected from Hund's rule.

In order to check the tendency of magnetism (ferromagnetic (FM)  or antiferromagnetic (AFM)) a pair of Fe ions are substituted into the cluster assuming a parallel (FM) as well as an anti-parallel (AFM) arrangement of magnetic moments and the structure of these clusters are relaxed.  The difference in total energies between the AFM and FM arrangement of magnetic moments ($\Delta E$) is displayed in Fig.~1(b) for seven possible configurations with increasing separation between Fe ions in the cluster. This energy is a measure of the inter-atomic exchange interaction. We observe that the coupling is always AFM and decays very fast with distance. This short-ranged AFM superexchange interaction has also been observed in the case of transition metal doped bulk ZnO.\cite{iusan06} This is due to the presence of transition metal states in the energy gap of the host cluster.  Our calculations reveal that it is not possible to stabilize ferromagnetism by doping Fe in a neutral charge state in pristine ZnO. However the abundance of room-temperature ferromagnetism in Mn, Fe and Co doped ZnO nanocrystals suggests that defects may play a crucial role to stabilize ferromagnetism in these systems. 
\begin{figure}
\includegraphics[scale=0.35]{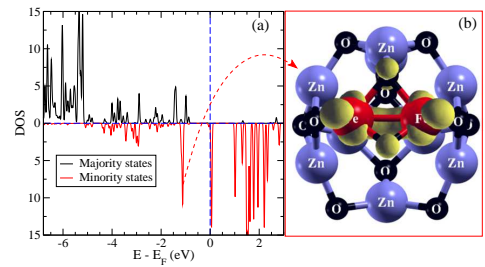}
\caption{\label{fig3}(Color online) (a): Partial DOS of Fe-d states and (b): charge density in an energy window corresponding to {\em one} minority spin state (indicated by a dashed arrow) in a cluster where a pair of  Fe atoms are doped at Zn sites near a Zn vacancy with charged state $-$1.}
\end{figure}

Therefore the formation energies were calculated for a single Fe atom doped into the cluster with and without defects (Zn and O vacancy) assumed in different charged states. We considered both Zn rich and O rich conditions.  The Fermi energy is varied from the  HOMO level of the pristine ZnO cluster to the value of the gap of bulk ZnO.  In Fig.~2(a) we display our results for the formation energy of Fe doped ZnO under oxygen rich conditions. From this figure we note that near the p-type region, Fe doped ZnO (without defects but in the charged state +1) has the lowest formation energy while in the n-type region, the lowest formation energy corresponds to the doped system with Zn vacancy in the charged state $-$1. The stability of the above two configurations provide an important evidence that the ionic state of the dopant Fe atom residing on the surface of the cluster may be Fe$^{3+}$ in agreement with recent experimental results\cite{karmakar} and may be attributed to the different co-ordination structure of the surface Fe atoms in comparison to the bulk terminated ones due to loss of ligands and co-ordination defects arising from lattice distortion.
In the Zn-rich limit the inclusion of defects does not lower the formation energy in either n-type or p-type region.

We have now examined whether the above two stable configurations may support the tendency for ferromagnetism. In Fig.~2(b),(c),(d) we have displayed the energy difference between the AFM and FM configuration by including two Fe atoms at the closest possible positions without and with  Zn and O vacancies in various charged states. We gather from the figure that for both of the above mentioned stable configurations there is a  tendency for ferromagnetism (See Fig.~2(b),(c)). The magnetic moment resides in each case predominantly on the dopant Fe atoms spilling over a bit onto the neighboring oxygen atoms. A pair of Fe atoms substituting Zn in the charged state $-$1 and +1 in the presence and absence of a Zn vacancy respectively point to the fact that the dopant Fe atoms are in the ionic states Fe$^{2+}$ ($d^{6}$) and Fe$^{3+}$ (d$^{5}$) (mixed valent). This ionic configuration is important to stabilize ferromagnetism as was speculated in a recent experiment.\cite{karmakar} This is further illustrated in Fig.~3 where we have considered a particular case namely a pair of Fe atoms substituted  into the ZnO cluster in the presence of a Zn vacancy in the charged state $-$1. For this case we plotted the Fe-d partial density of states (DOS) and show the charge density in a small energy window corresponding to {\em a single} Fe-d state in the minority spin channel as indicated in the figure. From the DOS we see that there is an admixture of the Fe-d states with the O-p states. Furthermore in the  majority spin channel both the Fe atoms are  completely occupied whereas in the minority spin channel only one Fe d state is occupied resulting in a net magnetic moment of 9 $\mu_{B}$.  The corresponding charge density for this Fe-d state in the minority spin channel reflects a hopping induced hybridization between the neighboring Fe ions which is responsible for promoting ferromagnetism by a reduction in energy in the process of hopping.  Although the oxygen vacancies do not have the lowest formation energies, their effect on the magnetic properties is also very interesting as it induces ferromagnetic interactions in otherwise antiferromagnetically coupled magnetic atoms in a defect-free system (See Fig.~2(d)).

In conclusion, we have studied the energetics and magnetic interactions in an Fe doped ZnO nano-cluster. It has been found that in a pure ZnO cluster the Fe atoms couple antiferromagnetically. The presence of Zn and O vacancies induces intra-cluster ferromagnetic coupling and hence a net magnetic moment for the cluster. We have argued that the ionic state of the dopant Fe ion plays a crucial role to make the cluster ferromagnetic.  Thus one can envisage the engineering of defects in nanoclusters to design spintronic devices.

ID thanks DST India for financial support. BS acknowledges the Swedish Research Council (VR) for financial support and Swedish National Infrastructure for Computing (SNIC) for granting computer time. ID and BS acknowledge Asia-Sweden Research Links Programme funded by VR/SIDA.



\end{document}